\input{aipcheck}

\documentclass[
    ,final            
  ]
  {aipproc}

\layoutstyle{6x9}

\begin{document}

\title{A Short Guide to Debris Disk Spectroscopy}


\author{Christine H. Chen}{
  address={Space Telescope Science Institute, 3700 San Martin Dr., Baltimore, MD 21218}
}

\begin{abstract}
Multi-wavelength spectroscopy can be used to constrain the dust and gas properties in debris disks. Circumstellar dust absorbs and scatters incident stellar light. The scattered light is sometimes resolved  spatially at visual and near-infrared wavelengths using high contrast imaging techniques that suppress light from the central star. The thermal emission is inferred from infrared through submillimeter excess emission that may be 1-2 orders of magnitude brighter than the stellar photosphere alone. If the  disk is not spatially resolved, then the radial distribution of the dust can be inferred from Spectral Energy Distribution (SED) modeling. If the grains are sufficiently small and warm, then their composition can be determined from mid-infrared spectroscopy. Otherwise, their composition may be determined from reflectance and/or far-infrared spectroscopy. Atomic and molecular gas absorb and resonantly scatter stellar light. Since the gas is believed to be secondary, detailed analysis analysis of the gas distribution, kinematics, and composition may also shed light on the dust composition and processing history. \end{abstract}

\maketitle


\section{Introduction}
Debris disks are dusty disks around main sequence stars that are distinguished from proto-planetary disks by their small gas:dust ratios. In the absence of bulk gas, the dynamics of the planets, small bodies, and dust grains can be calculated based on stellar (e.g. stellar luminosity and mass loss rate) and planetary (e.g. mass, orbital inclination, eccentricity, and semi-major axis) properties alone. Without gas to retard the loss of particles against radiation pressure or corpuscular stellar wind and Poynting-Robertson drag, circumstellar grains typically possess lifetimes of $<$10,000 years, significantly shorter than the age of the central star, implying that the grains are replenished from a reservoir. In these systems, unseen planets are presumed to perturb minor bodies such as asteroids or comets into crossing orbits, generating small dust grains that are detected via remote sensing. 

At the current time, radial velocity and transiting planet searches have discovered 291 companions to 251 solar-type stars with minimum masses $<$25 $M_{Jup}$ (see \url{http://exoplanets.eu}), 41 of which transit their host stars.  Statistical studies of their properties are expected to provide important constraints on planet formation models. For example, the observation that planets are found more frequently around metal-rich stars suggests that they typically form via core-accretion rather than gravitational instability \cite{fv05}. Debris disk studies can provide complementary constraints to planet formation and evolution models such as the bulk composition of exo-planets, the frequency and location of dust producing events generated during terrestrial planet formation and giant planet migration, and the final architecture of planetary systems. For example, the discovery of large masses of fine SiO$_{2}$ grains and Si gas around the $\beta$ Pic moving group member HR 7012 (with an age of $\sim$12 Myr) indicates that a recent massive collision has occurred in the inner planetary system, perhaps analogous to the large impact that may have stripped Mercury's mantle \cite{cl09}.

\section{Dust}
The most extensive spectroscopic observations of dust debris around main sequence stars have been carried out at mid-infrared wavelengths (5.5 - 35 $\mu$m) using the Infrared Spectrograph (IRS) on the \emph{Spitzer Space Telescope}. The majority of objects studied to date do not possess PAH or silicate emission features at 10 and/or 20 $\mu$m, suggesting that the dust grains are cold ($T_{gr}$ $\leq$ 110 K) and/or large ($2\pi a$ $>>$ $\lambda$) \cite{chc06}. Despite the lack of spectral features, the shape of the SED can be used to infer the spatial distribution of the dust in the absence of resolved images. However, such studies only provide approximate guidelines for the location of the dust because SED fitting is degenerate. Dust distances predicted from SED modeling, assuming single temperature black bodies, may be as much as a factor of two smaller than resolved disk sizes.

\subsection{Constraining the Spatial Distribution}
The SEDs of the majority of debris disks are well fit using single temperature black bodies suggesting that the dust in these systems is located in rings \cite{chc06}. As many as one-third of debris disk spectra may be better fit using multiple temperature components indicating either the presence of multiple components or a continuous disk with radii that extend several 10's of AU \cite{lah08}. The presence of central clearings may suggest that planets have already formed and are sculpting the disks \cite{fq07}; however, other processes that do not require the presence of planets may also produce the inferred central clearings such as  (1) radiation pressure if the disk is collisionally dominated and the grains rapidly shatter to sizes below the blow-out size, (2) sublimation if the grains are icy, and (3) grain sorting via gas drag if the disk has a dust:gas ratio of 0.1-10 \cite{ta01}. Estimates for the grain lifetimes of \emph{IRAS}-discovered debris disks (that are better fit using single temperature black bodies) suggest that these systems are collision dominated and that their central clearings are generated by radiation pressure \cite{chc06}.

The dust in our solar system is more tenuous and therefore spirals in under corpuscular solar wind and Poynting-Robertson (CPR) drag before it collides with other dust. If dust in debris disks, produced in collisions between parent bodies, spirals in under CPR drag, then it is expected to generate a disk with a uniform surface density and a flux,
\begin{equation}
F_{\nu} \propto \nu^{-1} \label{powerlawsed}
\end{equation}
\cite{jura98} independent of the mass opacity wavelength dependence (i.e. for any $\kappa_{\nu}$ $\propto$ $\nu^{p}$). Collisional cascade studies suggest that the boundary between the collisionally- and PR-drag dominated regimes around main sequence A-tye stars should occur at fractional luminosities, $L_{IR}/L_{*}$ $\sim$ 10$^{-4}$ \cite{kmk00}; however, all of the \emph{IRAS}-discovered debris disks appear to be collisionally-dominated even though some objects possess  $L_{IR}/L_{*}$ as small as 10$^{-6}$. $L_{IR}/L_{*}$ may not be the best metric to determine whether disks are collisionally- or CPR dominated because this quantity is proportional to dust mass rather than dust density. Recently, a dozen dusty disks around main sequence A-type stars have been discovered to possess power-law SEDs \cite{fym09}. Since the estimated PR-drag lifetime of the observed dust is smaller than the collisional lifetime, the dust in these systems may be PR-drag dominated.  

This past year has provided some of the most stunning confirmation that exo-planets and debris disks are intricately intertwined. The first direct detections of orbiting exo-planets were made in debris disk systems: (a) thermal emission from three $\sim$10 M$_{Jup}$ planets around HR 8799 \cite{cm08} and (b) scattered light from a cirumplanetary disk around a <5 M$_{Jup}$ planet in the Fomalhaut disk \cite{pk08}. SED modeling of HR 8799 indicates that this system may be a solar system analog with both asteroidal dust interior to and Kuiper-belt dust exterior to the orbits of the three newly discovered planets. Minimum $\chi^{2}$ fitting of the IRS excess at wavelengths shorter than 30 $\mu$m indicates the presence of large, warm dust grains with $T_{gr}$ = 160 K, located at a distance $\sim$8 AU. The remaining IRAS 60 $\mu$m and JCMT SCUBA 850 $\mu$m excesses indicate the presence of small, cold dust grains with an emissivity, $\kappa_{\nu}$ $\propto$ $\nu$, and $T_{gr}$ = 40 K, located at a distance of 2000 AU.

\subsection{Constraining the Grain Composition}
Dust produced in large collisions during the late-stages of terrestrial planet formation ($<$100 Myr), the period of Late Heavy Bombardment ($\sim$700 Myr), and in steady state grinding of asteroid belts is sufficiently warm that any small silicate grains present are expected to produce mid-infrared emission features. Such features have been discovered and/or exquisitely characterized toward a couple dozen, typically young ($<$50 Myr) debris disks using \emph{Spitzer} IRS. The majority of the systems with silicate emission features possess both warm crystalline silicates and cool black body components, indicative of the presence of multiple parent body belts \cite{chc06}. Early paradigms for dust processing in disks envisioned the initial incorporation of small, amorphous silicates from the ISM in proto-planetary disks that are subsequently annealed into large, crystalline grains, analogous to those found in comets; however, observations of dust around stars of similar age and spectral type reveal a diversity of composition and processing history. For example, the mid-infrared spectrum of A6V star $\beta$ Pic possesses olivine and forsterite features \cite{chc07} while the spectra of A0V $\beta$ Pic moving group members HR 7012 and $\eta$ Tel possess strong enstatite and no crystalline features, respectively \cite{chc06}.
 
 Since the majority of systems possess cold dust, alternative techniques are necessary to determine their grain compositions. Currently, 20 debris disks have been spatially resolved with coronagraphic imaging at visual and near-infrared wavelengths (see \url{http://astro.berkeley.edu/~kalas/disksite/pages/gallery.html}). Broad-band visual and near-infrared reflectance photometry and low-resolution spectroscopy has been proposed as a technique to distinguish water ice grains from organic and silicate grains. Low resolution \emph{HST} STIS visual spectroscopy (R $\sim$ 1000) of the reflected light around from the disk around TW Hya revealed gray scattering, suggesting the presence of large grains ($a$ $>$ 1 $\mu$m) \cite{aki05}. Broad-band \emph{HST} ACS, STIS, and NICMOS scattered light photometry of the reflected light from the disk around around HR 4796A revealed red scattering coefficients, suggesting the presence of tholins \cite{dws08} and/or silicates \cite{kml08}; however, broad-band photometry does not possess sufficient spectral resolution to select among models. In the future, modest resolution reflectance spectroscopy at both visual and near-infrared wavelengths may be able to determine uniquely the composition of the grains by measuring not only their scattered light color but also ice features.
 
Since the bulk of the thermal emission for cold black bodies (with $T_{gr}$ $<$ 100 K) emerges at far-infrared wavelengths ($\lambda$ $\sim$ 70-100 $\mu$m), far-infrared spectroscopy may also provide insight into grain composition. ISO provided moderate resolution spectroscopy (R $\sim$1000) for the brightest objects in the far-infrared sky with fluxes, $F_{\nu}$(70 $\mu$m) $\sim$ 100 Jy, significantly higher than the typical fluxes measured toward debris disks, $<$1 Jy. The ISO spectrum of the pre-main sequence F7IIIe star HD 142527 revealed not only a hot dust component (500 - 1500 K) containing silicates and some C-rich dust (graphite and [CII]) but also a cold dust component (30 - 60 K) containing O-rich dust and some crystalline water ice and hydrous silicates \cite{km99} . Resolved scattered-light and thermal emission images of the dust around HD 181327 have been recently modeled self-consistently with the \emph{Spitzer} IRS and MIPS SED-mode spectra (with resolutions of R $\sim$ 100 and 10, respectively) assuming that the dust possesses 75\% amorphous silicates and 25\% water ice, somewhat less than the 60\% water ice observed toward HD 142527 \cite{chc08}.

\section{Gas}
Core accretion models that describe the formation of our solar system estimate that the Jovian planet cores coagulated on a timescale of $\sim$1 Myr and accreted their gas-rich envelopes on a timescale $\sim$10-30 Myr. Searches for bulk H$_{2}$ and CO around  main sequence stars have been made with the aim of detecting the gas accretion phase of giant planet formation. Since the gas was expected to be cool ($T_{ex}$ $\sim$ 100 K), \emph{Spitzer} IRS has been used to search for emission from the S(1) and S(0) transitions of H$_{2}$ at 17.1 $\mu$m and 28.0 $\mu$m, respectively, and submilleter heterodyne receivers have been used to search for emission from the J = 2$\rightarrow$1  and J = 3$\rightarrow$2 transitions of $^{12}$CO at 230 GHz and 345 GHz, respectively; however, no bulk gas has been detected around the majority of young main sequence stars \cite{ilaria06, chc07}. Although debris disks possess too little gas to form giant planets, they may possess sufficient gas to impact the dynamics of small grains. More sensitive ultra-violet and visual absorption line studies have detected circumstellar gas in edge-on debris disk around late B- and early A-type stars.

Ultra violet and visual spectra of $\beta$ Pictoris serendipitously discovered the presence of time-variable, high velocity, red-shifted absorption features, initially in Ca II H and K and Na I D and later in a suite of atoms (including C I, C IV, Mg I, Mg II, Al II, Al III, Si II, S I, Cr II, Fe I, Fe II, Ni II, Zn II); these features vary on timescales as short as hours and are non-periodic \cite{vlf98}. Since the velocity of the atoms, typically 100 km/sec - 400 km/sec, is close to the free fall velocity at a few stellar radii, the absorption is believed to be produced in the comas of infalling refractory bodies at distances $<$6 AU from the star \cite{kbk01, beu98}. At these distances, refractory materials sublimate from the surface of infalling bodies and collisions produce highly ionized species such as C IV and Al III. If infalling bodies generate the observed features, then the fact that the features are preferentially red-shifted (rather than equally red- and blue-shifted) may suggest that an unidentified process aligns their orbits. Scattering of bodies by a planet on an eccentric orbit \cite{ldw94} and orbital decay via secular resonance perturbations \cite{bm96} have been used to explain the preferentially red-shifted features; however, both models have difficulties.

The $\beta$ Pictoris disk also possesses a stable gas component at the velocity of the star. Spatially resolved visual spectra of $\beta$ Pic have revealed the presence of a rotating disk of atomic gas, observed via resonantly scattered emission from Fe I, Na I, Ca II, Ni I, Ni II, Ti I, Ti II, Cr I, and Cr II. Estimates of the radiation pressure acting on Fe and Na atoms suggest that these species should be accelerated to terminal velocities $\sim$100s - 1000s km/sec, significantly higher than observed \cite{alexis04}. The outward flow of gas may be retarded if the gas is overabundant in carbon because carbon does not possess strong far-UV resonance lines and can couple to other species via the Coulomb force \cite{fbw06}. \emph{FUSE} measurements suggest that carbon is overabundant in the $\beta$ Pic disk by a factor of 10-20, compared with measurements of other species from the literature \cite{aki06}. The origin of the stable atomic gas component is currently not well understood. Infalling refractory bodies undoubtably contribute at least some material \cite{bv07}. The similarity in the spatial distribution of the dust and gas suggests that the gas may be produced from the dust. One possibility is that sub-blow out sized grains on radial trajectories collide with larger orbiting particles with high relative velocities, directly vaporizing grain material \cite{cm07}. Another possibility is that far-UV stellar photons photo-desorb atoms from the surface of dust grains \cite{chc07}; however, neither of these possibilities correctly reproduces the spatial distribution of the gas.

\section{Future Work}
The \emph{Spitzer} IRS has opened a gateway to studying the structure and composition of debris disks using mid-infrared spectroscopy. Four large IRS debris disks studies have been published thus far \cite{cb06, chc06, lah08, fym09} with many more anticipated. Although low resolution mid-infrared spectra characterize the Wien portion of the dust thermal SED well, complimentary low resolution far-infrared spectra are needed to characterize the Rayleigh-Jeans portion and to complete the SED. Although \emph{Herschel} will launch in May 2009 and obtain images and spectra of the far-infrared sky, it is not sensitive enough to measure the $\sim$50 - 500 $\mu$m SED with R$\sim$100 for a large sample of debris disks. Future far-infrared observatories are still needed. 

The \emph{James Webb Space Telescope} is currently expected to launch in 2013 and to provide near- and mid-infrared spectroscopy and coronagraphic imaging of dozens to hundreds of new disks with its 6.5 m multi-segmented mirror. In particular, the Mid-Infrared Instrument (MIRI)  will provide an image slicer with spectral resolution R=1200-2400 at 5-29 $\mu$m to obtain spatially resolved spectra of extended disks. Its unprecedented sensitivity will also provide unresolved spectra of more distant objects. The Tunable Filter Imager (TFI) will provide R$\sim$100 coronagraphic imaging at 1.6-4.9 $\mu$m allowing it to search for spectral signatures due to ice and organics in reflected light.

Ground based visual spectroscopic studies searching for and characterizing circumstellar atomic gas are on-going. The \emph{Hubble Space Telescope} Servicing Mission 4 (May 2009) is expected to restore UV absorption-line spectroscopy (COS, STIS) to measure the composition and kinematics of edge-on gas disks and to restore high contrast coronagraphic imaging (ACS, STIS, and NICMOS) to measure the reflectance properties of dusty disks. The Gas in Protoplanetary Systems (GASPS) key programme has been awarded 400 hours of telescope time to conduct a survey searching for emission from [C II] and [O I] at 157 $\mu$m and 63 $\mu$m around 274 stars with ages 1 - 30 Myr. Undoubtably, many new discoveries will be made in the coming decade that will broaden our understanding of the evolution of debris disks and their attendant planets from measurements of the spatial distribution and composition of their circumstellar material.

\end{document}